\documentclass[twocolumn,twoside]{article}
\usepackage{graphicx,aasab}
\usepackage[super,sort&compress]{natbib}
\renewcommand{\baselinestretch}{0.949}

\bibpunct[,]{(}{)}{,}{s}{ }{,}

\newcommand{\sgrastar}{Sgr~A\mbox{$^*$}}

\begin{document}

\markright{Nature, Vol.~413, pp.~45--48 (2001)}

\begin{flushleft}
{\raggedright\Large\bf\noindent Rapid X-ray flaring from the direction
of the supermassive black hole at the Galactic Centre}
\end{flushleft}

\begin{flushleft}
{\bf\noindent F.~K.~Baganoff$^\star$, M.~W.~Bautz$^\star$,
W.~N.~Brandt$\dagger$, G.~Chartas$\dagger$, E.~D.~Feigelson$\dagger$,
G.~P.~Garmire$\dagger$, Y.~Maeda$\dagger\ddagger$, M.~Morris\S,
G.~R.~Ricker$^\star$, L.~K.~Townsley$\dagger$ \& F.~Walter$\parallel$}
\end{flushleft}

\begin{flushleft}
\footnotesize\noindent
\emph{$^\star$ Center for Space Research, Massachusetts Institute of
  Technology, Cambridge, MA 02139-4307, USA}\\
\emph{$\dagger$ Department of Astronomy and Astrophysics, Pennsylvania
  State University, University Park, PA 16802-6305, USA}\\
\emph{$\ddagger$ Institute of Space and Astronautical Science, 3-1-1
  Yoshinodai, Sagamihara, 229-8501, Japan}\\
\emph{\S Department of Physics and Astronomy, University of California
  at Los Angeles, Los Angeles, CA 90095-1562, USA}\\
\emph{$\parallel$ Department of Astronomy, California Institute of
  Technology, Pasadena, CA 91125, USA}
\end{flushleft}

\medskip

{\small

{\bf\noindent Most galactic nuclei are now believed to harbour
supermassive black holes\nolinebreak\cite{Richstone98}.  Studies of
stellar motions in the central few light-years of our Milky Way Galaxy
indicate the presence of a dark object with a mass of
$\mathbf{\approx2.6 \times 10^6}$ solar masses (refs 2, 3).  This
object is spatially coincident with Sagittarius~A$^*$ (\sgrastar), the
unique compact radio source located at the dynamical centre of our
Galaxy.  By analogy with distant quasars and nearby active galactic
nuclei (AGN), \sgrastar\ is thought to be powered by the gravitational
potential energy released by matter as it accretes onto a supermassive
black hole\nolinebreak\cite{Lynden-Bell71, Melia01c}.  However,
\sgrastar\ is much fainter than expected in all wavebands, especially
in X-rays, casting some doubt on this model.  Recently, we reported
the first strong evidence of X-ray emission from \sgrastar\ (ref.~6).
Here we report the discovery of rapid X-ray flaring from the direction
of \sgrastar.  These data provide compelling evidence that the X-ray
emission is coming from accretion onto a supermassive black hole at
the Galactic Centre, and the nature of the variations provides strong
constraints on the astrophysical processes near the event horizon of
the black hole.}

Our view of \sgrastar\ in the optical and ultraviolet wavebands is
blocked by the large visual extinction, $A_{\rm V} \approx
30$~magnitudes\nolinebreak\cite{Morris96}, caused by dust and gas
along the line of sight.  \sgrastar\ has not been detected in the
infrared due to its faintness and to the bright infrared background
from stars and clouds of dust\nolinebreak\cite{Menten97}.  Detection
of X-rays from \sgrastar\ is therefore essential to constrain the
spectrum at energies above the radio-to-submillimetre band and to test
the supermassive-black-hole accretion-flow
paradigm\nolinebreak\cite{Melia01c}.

We first observed the Galactic Centre on 21 September 1999 with the
imaging array of the Advanced CCD Imaging Spectrometer (ACIS-I) aboard
the \emph{Chandra X-ray Observatory}\nolinebreak\cite{Weisskopf96} and
discovered an X-ray source coincident within $0\farcs35 \pm 0\farcs26$
($1\sigma$) of the radio source\nolinebreak\cite{Baganoff01}.  The
luminosity in 1999 was very weak, $L_{\rm X} \approx 2 \times 10^{33}$
erg s$^{-1}$ in the 2--10~keV band, after correction for the inferred
neutral hydrogen absorption column $N_{\rm H} \approx 1 \times
10^{23}$ cm$^{-2}$.  This is far fainter than previous X-ray
observatories could detect\nolinebreak\cite{Baganoff01}.

We observed the Galactic Centre a second time with
\emph{Chandra}/ACIS-I from 26 October 2000 22:29 through 27 October
2000 08:19 (UT), during which time we saw a source at the position of
\sgrastar\ brighten dramatically for a period of $\approx10$~ks.
Figure~\ref{fig:surfplot} shows surface plots for both epochs of the
2--8~keV counts integrated over time from a $20\arcsec \times
20\arcsec$ region centred on the radio position of \sgrastar.  The
modest peak of the integrated counts at \sgrastar\ in 1999 increased
by a factor of $\approx7$ in 2000, despite the 12\% shorter exposure.
The peak integrated counts of the fainter features in the field show
no evidence of strong variability, demonstrating that the flaring at
\sgrastar\ is intrinsic to the source.

\begin{figure}[!htb]
\begin{center}
\includegraphics[width=0.484\textwidth]{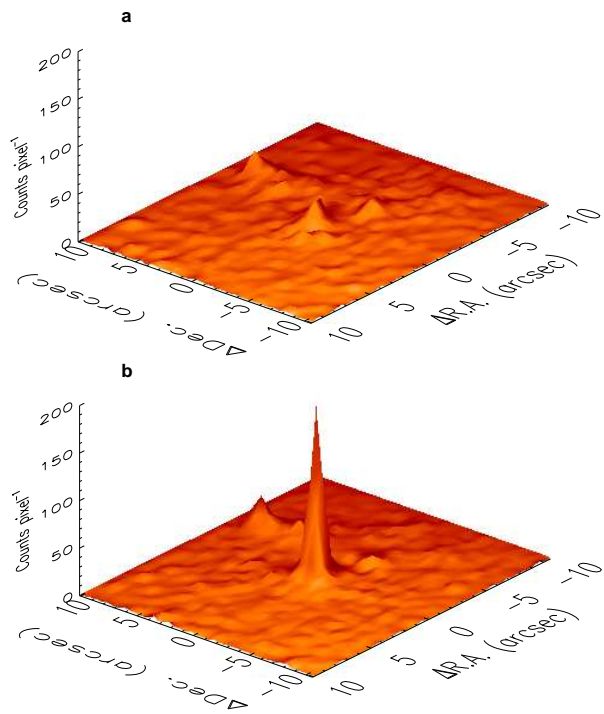}
\caption{\footnotesize Surface plots of the 2--8~keV counts within a
$20\arcsec \times 20\arcsec$ field centred on \sgrastar\ at two
epochs.  The data were taken with \emph{Chandra}/ACIS-I on ({\bf a})
21 September 1999 and ({\bf b}) 26--27 October 2000.  The effective
exposure times were 40.3~ks and 35.4~ks, respectively.  The spatial
resolution is 0\farcs5 per pixel.  An angle of 1\arcsec\ on the sky
subtends a projected distance $\approx0.04$~pc at the galactocentric
distance of 8.0~kpc (ref.~30).  The peak integrated counts per pixel
at the position of \sgrastar\ increased by a factor of $\approx7$ from
the first epoch to the second, despite the slightly smaller exposure
time ($\approx12\%$) in the second epoch.  The low-level peak a few
arcseconds to the southwest of \sgrastar\ is the infrared source
IRS~13, while the ridge of emission to the northwest is from a string
of unresolved point sources.  The fainter features in the field are
reasonably consistent between the two epochs, considering the limited
Poisson statistics and the fact that these stellar sources may
themselves be variable; this consistency shows that the strong
variations at \sgrastar\ are intrinsic to the source.}
\label{fig:surfplot}
\end{center}
\end{figure}

Figure~\ref{fig:ltc} shows light curves of the photon arrival times
from the direction of \sgrastar\ during the observation in 2000.
Panels (a) and (b) show hard-band (4.5--8~keV) and soft-band
(2--4.5~keV) light curves constructed from counts within an angular
radius of 1\farcs5 of \sgrastar.  Both bands exhibit roughly constant,
low-level emission for the first $\approx14$~ks, followed by an
$\approx6$-ks period of enhanced emission beginning with a 500-s event
($4.4\sigma$ significance using 150-s bins).  At 20~ks, there occurs a
large-relative-amplitude flare or flares lasting $\approx10$~ks, and
finally a return to the low state for the remaining $\approx6$~ks.
About 26~ks into the observation, the hard-band light curve drops
abruptly by a factor of $\approx5$ within a span of $\approx600$~s and
then partially recovers within a period of $\approx1.2$~ks.  The
soft-band light curve shows a similar feature, but it appears to lag
the hard-band event by a few hundred seconds and is less sharply
defined.

\begin{figure}[!htb]
\begin{center}
\includegraphics[width=0.484\textwidth]{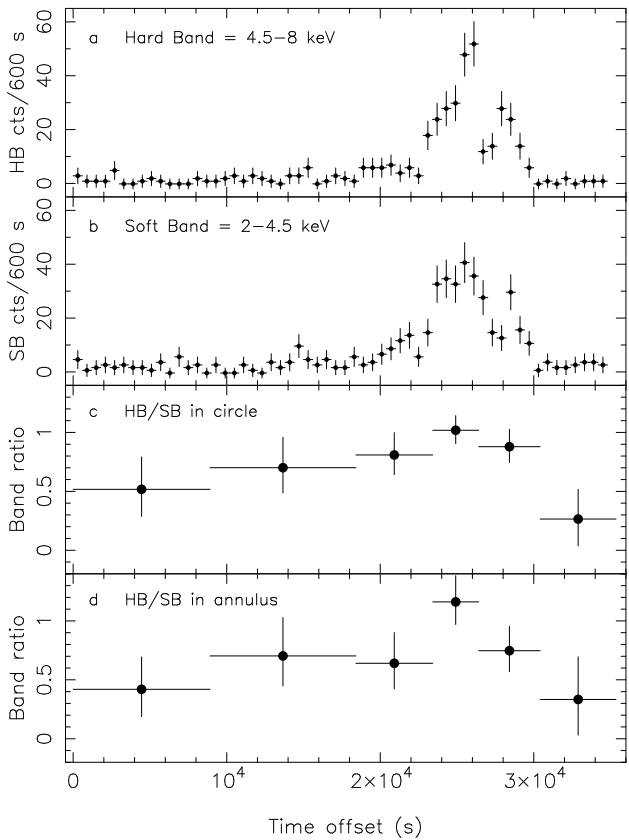}
\caption{\footnotesize Light curves of the photon arrival times and
band ratios from the direction of \sgrastar\ on 26--27 October 2000.
Panels ({\bf a}) hard-band (HB = 4.5--8~keV) counts, ({\bf b})
soft-band (SB = 2--4.5~keV) counts, ({\bf c}) hard/soft band ratio
within a circle of radius 1\farcs5, and ({\bf d}) hard/soft band ratio
within an annulus of inner and outer radii 0\farcs5 and 2\farcs5,
respectively.  The x-axis shows the time offset from the start of the
observation at 26 October 2000 22:29 (UT).  The data are shown with
$1\sigma$ error bars.  The single 2.8-hour period of flaring activity
which we have detected so far during a total of 21 hours of
observations yields a (poorly determined) duty cycle of $\sim1/8$.
Our continuing observations of this source with \emph{Chandra} will
permit us to refine this value.  During the quiescent intervals at the
beginning and end of this observation, the mean count rate in the
2--8~keV band was $(6.4 \pm 0.6) \times 10^{-3}$ counts s$^{-1}$,
consistent with the count rate we measured in 1999, which was $(5.4
\pm 0.4) \times 10^{-3}$ counts s$^{-1}$.  The detected count rate at
the peak of the flare was $0.16 \pm 0.01$ counts s$^{-1}$ within a
1\farcs5-radius extraction circle, but this is $\approx30$\% less than
the true incident count rate due to pile-up of X-rays in the detector
during the 3.2-s integration time per CCD read out.}
\label{fig:ltc}
\end{center}
\end{figure}

The band-ratio time series in panel (c), defined as the ratio of
hard-band counts to soft-band counts, suggests the spectrum hardened
during the flare.  The difference between the band ratio measured at
the peak of the flare and the average of the band ratios during the
quiescent periods at the beginning and end of the observation is $0.63
\pm 0.21$ (i.e., $3\sigma$).  The peak-flare band ratio in panel (c)
is affected to some extent by the effects of pile-up (see
Figure~\ref{fig:ltc} caption), which would tend to harden the
spectrum; however, the band ratios in panel (d), which were computed
using the non-piled-up data extracted from the wings of the point
spread function, also show evidence for spectral hardening with
$2.7\sigma$ significance.  The sizes of dust-scattering halos in the
Galactic Centre are typically $\ga1\arcmin$ (ref.~10), so
dust-scattered X-rays from the source contribute a negligible fraction
of the emission within the source extraction region that we used;
hence it cannot account for the spectral variations.  We therefore
conclude that the spectral hardening during the flare is likely to be
real.

\markboth{Baganoff et~al.---Discovery of rapid X-ray flaring from the
direction of \sgrastar}{Baganoff et~al.---Discovery of rapid X-ray
flaring from the direction of \sgrastar}

The quiescent-state spectra in 1999 and 2000 and the peak
flaring-state spectrum in 2000 are shown in Figure~\ref{fig:spectra}.
We fit each spectrum individually using a single power-law model with
corrections for the effects of photoelectric absorption and dust
scattering\nolinebreak\cite{Predehl95}.  The best-fit values and 90\%
confidence limits for the parameters of each fit are presented in the
first three lines of Table~\ref{tab:fits}.  The column densities for
the three spectra are consistent, within the uncertainties, as are the
photon indices of both quiescent-state spectra.  Next, we fit a double
power-law model to the three spectra simultaneously, using a single
photon index for both quiescent spectra, a second photon index for the
flaring spectrum, and a single column density for all three spectra.
The best-fit models for each spectrum from the simultaneous fits are
shown as solid lines in Figure~\ref{fig:spectra}; the parameter values
are given in the last line of Table~\ref{tab:fits}.  Using these
values, we derive an absorption-corrected 2--10~keV luminosity of
$L_{\rm X} = (2.2_{-0.3}^{+0.4}) \times 10^{33}$ erg s$^{-1}$ for the
quiescent-state emission and $L_{\rm X} = (1.0 \pm 0.1) \times
10^{35}$ erg s$^{-1}$ for the peak of the flaring-state emission, or
$\approx45$ times the quiescent-state luminosity.  We note that
previous X-ray observatories did not have the sensitivity to detect
such a short-duration, low-luminosity flare in the Galactic Centre
\nolinebreak\cite{Baganoff01}.  The best-fit photon index $\Gamma_{\rm
f} = 1.3_{-0.6}^{+0.5}$ ($N(E) \propto E^{-\Gamma}$) for the
flaring-state spectrum is slightly flatter than, but consistent with,
systems thought to contain supermassive black
holes\nolinebreak\cite{Mushotzky93}.

\begin{figure}[!htb]
\begin{center}
\includegraphics[width=0.484\textwidth]{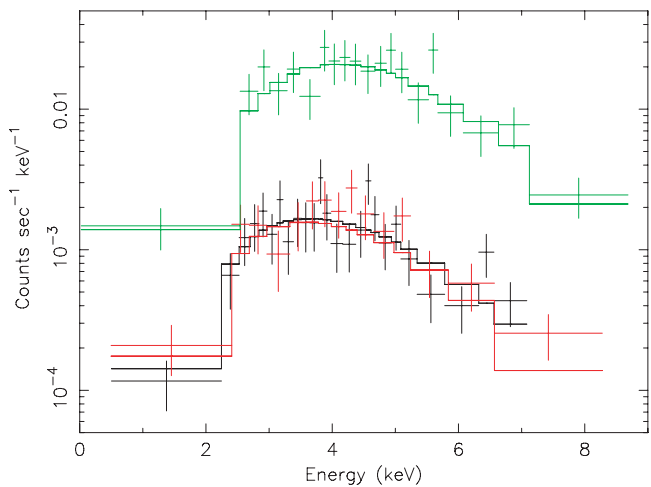}
\caption{\footnotesize X-ray spectra of the \emph{Chandra} source at
the position of \sgrastar.  The data are shown as crosses with
vertical bars indicating the $1\sigma$ errors in the count rate and
horizontal bars the energy range of each bin.  The events have been
grouped to yield 10 counts per bin.  The counts in the 1999 (black)
and 2000 (red) quiescent-state spectra were extracted using a source
radius of 1\farcs5.  A non-piled-up, peak flaring-state spectrum
(green) was extracted from the wings of the point spread function,
using an annulus with inner and outer radii of 0\farcs5 and 2\farcs5,
during the time interval 23.7--26.3~ks after the start of the 2000
observation.  The solid lines are the best-fit models for each
spectrum, obtained by fitting an absorbed, dust-scattered, double
power-law model to the three spectra simultaneously.  The best-fit
values and 90\% confidence intervals for the model parameters are
given in the last line of Table~\ref{tab:fits}.  The spectra are
well-fitted using a single photon index for both quiescent spectra, a
second photon index for the flaring spectrum, and a single column
density for all three spectra.  The column density from the
simultaneous fits corresponds to a visual extinction $A_{\rm V} =
29.6_{-6.1}^{+5.0}$~magnitudes, which agrees well with
infrared-derived estimates of $A_{\rm V} \approx
30$~magnitudes\protect\nolinebreak\cite{Morris96}; we thus find no
evidence in our X-ray data of excess gas and dust localized around the
supermassive black hole.  This places an important constraint on the
maximum contribution to the infrared spectrum of \sgrastar\ produced
by local dust reprocessing of higher energy photons from the accretion
flow.  The spectral models used in ref.~6 did not account for dust
scattering; hence a higher column density was needed to reproduce the
low-energy cut-off via photoelectric absorption alone.  We note there
is no sign of an iron K$\alpha$ emission line in the flaring-state
spectrum.}
\label{fig:spectra}
\end{center}
\end{figure}

\begin{table}[!hb]
\begin{center}
\footnotesize
\caption{Spectral Fits}
\label{tab:fits}
\begin{tabular}[0.484\textwidth]{lcccc}
\multicolumn{5}{l}{} \\
\hline
\multicolumn{5}{l}{} \\
\multicolumn{1}{c}{Spectrum} &
\multicolumn{1}{c}{$N_{\rm H}$} &
\multicolumn{1}{c}{$\Gamma_{\rm q}$} &
\multicolumn{1}{c}{$\Gamma_{\rm f}$} &
\multicolumn{1}{c}{$\chi^2/\nu$} \\
\multicolumn{5}{l}{} \\
\hline
\multicolumn{5}{l}{} \\
 1999 Quiescent  & $5.8_{-1.4}^{+1.3}$ & $2.5_{-0.7}^{+0.8}$ & \nodata             & \phd19/22 \\[1ex]
 2000 Quiescent  & $5.0_{-2.3}^{+1.9}$ & $1.8_{-0.9}^{+0.7}$ & \nodata             &    7.6/12 \\[1ex]
 2000 Flaring    & $4.6_{-1.6}^{+2.0}$ & \nodata             & $1.0_{-0.7}^{+0.8}$ & \phd12/17 \\[1ex]
 All             & $5.3_{-1.1}^{+0.9}$ & $2.2_{-0.7}^{+0.5}$ & $1.3_{-0.6}^{+0.5}$ & \phd45/55 \\
\multicolumn{5}{l}{} \\
\hline
\end{tabular}
\end{center}
\parbox{0.484\textwidth}{
\scriptsize
\renewcommand{\baselinestretch}{0.7}
\noindent Best-fit parameter values and 90\% confidence intervals for
power-law models, corrected for photoelectric absorption and dust
scattering\protect\nolinebreak\cite{Predehl95}.  $N_{\rm H}$ is the
neutral hydrogen absorption column in units of $10^{22}$ H~atoms
cm$^{-2}$.  $\Gamma_{\rm q}$ and $\Gamma_{\rm f}$ are the
photon-number indices of the quiescent-state and peak flaring-state
spectra ($N(E) \propto E^{-\Gamma}$).  $\chi^2$ is the value of the
fit statistic for the best-fit model and $\nu$ is the number of
degrees of freedom in the fit.  The parameter values for the spectrum
marked ``All'' were derived by fitting an absorbed, dust-scattered,
double power-law model to the three spectra simultaneosly (see text
and Figure~\ref{fig:spectra}).}
\end{table}

If we view the outburst as a single event, the few-hundred-second
rise/fall timescales and the 10-ks duration are consistent with the
light-crossing and dynamical timescales for the inner $\la10$
Schwarzchild radii ($R_{\rm s} \equiv 2GM/c^2$) of the accretion flow
around a black hole of $2.6 \times 10^{6}$ solar masses; here $R_{\rm
s}$ is the radius of the black-hole event horizon (i.e., the boundary
at which the escape velocity equals the speed of light), $G$ is the
gravitational constant, $M$ is the mass of the black hole, and $c$ is
the speed of light.  While we cannot strictly rule out an unrelated
contaminating source as the origin of the flare (e.g., an X-ray
binary, for which little is known about such short-timescale,
low-luminosity events as we have detected; W.~Lewin, private
communication), this explanation seems unlikely since the
characteristic angular scales of the young and old stellar clusters
around \sgrastar\ are 5--20\arcsec\ (ref.~7), whereas the flaring
source lies within 1/3\arcsec\ of the radio position.  These clusters
contain up to a million-solar-masses worth of stars and stellar
remnants\nolinebreak\cite{Genzel00}; hence it is rather improbable
that there would be only one very unusual stellar X-ray source in the
image and that it would be fortuitously superposed on \sgrastar.
Furthermore, it is not clear that X-ray binaries can be easily formed
or long endure near \sgrastar, given the high velocity dispersion and
high spatial density of the stars in its deep gravitational potential
well\nolinebreak\cite{Davies98, Baganoff01}.

Strong, variable X-ray emission is a characteristic property of AGN;
factors of $\sim2$--3 variations on timescales ranging from minutes to
years are typical for radio-quiet AGN\nolinebreak\cite{Mushotzky93}.
Moderate- to high-luminosity AGN (i.e., Seyfert galaxies and quasars)
show a general trend of increasing variability with decreasing
luminosity\nolinebreak\cite{Nandra97}.  However, this trend does not
extend to low-luminosity active galactic nuclei (LLAGN), which show
little or no significant variability on timescales less than a
day\nolinebreak\cite{Ptak98}.  Assuming the X-ray flare is from
\sgrastar, it is remarkable that this source---generally thought to be
the nearest and least luminous example of accretion onto a central
supermassive black hole---has shown a factor of $\approx45$ variation
that is an order of magnitude more rapid than the fastest observed
variation of similar relative amplitude by a radio-quiet AGN of any
luminosity class\nolinebreak\cite{Ulrich97}.  We note that flares of
similar luminosity would be undetectable by \emph{Chandra} in the
nucleus of even the nearest spiral galaxy, M31.  LLAGN emit $L_{\rm X}
\ga 10^{38}$ erg s$^{-1}$ (ref.~14), so it should be kept in mind
that the astrophysics of accretion onto even the LLAGN may differ
substantially from that of \sgrastar.  This makes \sgrastar\ a
uniquely valuable source for testing the theory of accretion onto
supermassive black holes in galactic nuclei.

The faintness of \sgrastar\ at all wavelengths requires that the
supermassive black hole be in an extremely quiet phase, either because
the accretion rate is very low, or because the accretion flow is
radiatively inefficient, or both\nolinebreak\cite{Melia01c}.  A
variety of theoretical scenarios, usually based on advective accretion
models\nolinebreak\cite{Narayan98, Quataert99, Ball01, Blandford99},
jet-disk models\nolinebreak\cite{Falcke00}, or Bondi-Hoyle
models\nolinebreak\cite{Melia94, Melia01a}, have developed this idea.
An important prediction of the advective accretion models is that the
X-ray spectrum in the \emph{Chandra} energy band should be dominated
by thermal bremsstrahlung emission from hot gas in the outer regions
of the accretion flow ($R \ga 10^3\ R_{\rm s}$), but a region this
large could not produce the rapid, large-relative-amplitude variations
we have seen.  Thus, the properties of the X-ray flare are
inconsistent with the advective accretion flow models.  The low
luminosity and short timescales of this event are also inconsistent
with tidal disruption of a star by a central supermassive black
hole\nolinebreak\cite{Rees88}.

In all models, the radio-to-submillimetre spectrum of \sgrastar\ is
cyclo-synchrotron emission from a combination of sub-relativistic and
relativistic electrons (and perhaps positrons) spiralling around
magnetic field lines either in a jet or in a static region within the
inner $10\ R_{\rm s}$ of the accretion flow.  The electron Lorentz
factor inferred from the radio spectrum of \sgrastar\ is $\gamma_e
\sim 10^2$.  If the X-ray flare were produced via direct synchrotron
emission, then the emitting electrons would need $\gamma_e \ga 10^5$.
For the 10--100 G magnetic field strengths predicted by the models,
the cooling time of the particles would be $\sim1$--100~s.  Thus, the
$\approx10$-ks duration of the flare would require repeated injection
of energy to the electrons.  On the other hand, if the X-rays were
produced via up-scattering of the submillimetre photons off of the
relativistic electrons, a process called synchrotron
self-Comptonization (SSC), then $\gamma_e \sim 10^2$--$10^3$ would be
required, and the cooling time would be of order hours, which is
consistent with the duration of the flare.  The rapid turn-off of the
X-ray emission might then be attributed to the dilution of both photon
and electron densities in an expanding plasma.

The X-ray spectra of radio-quiet quasars and AGN are thought to be
produced by thermal Comptonization of infrared-to-ultraviolet seed
photons from a cold, optically thick, geometrically thin accretion
disk by hot electrons in a patchy corona above the
disk\nolinebreak\cite{Haardt97, Nandra97}.  The X-ray spectra of these
sources generally soften as they brighten\nolinebreak\cite{Ulrich97}.
In contrast, the extremely low luminosity of \sgrastar\ precludes the
presence of a standard, optically thick accretion
disk\nolinebreak\cite{Melia01c}; hence, the dominant source of seed
photons would be the millimetre-to-submillimetre synchrotron photons.

The energy released by an instability in the mass accretion rate or by
a magnetic reconnection event near the black hole would shock
accelerate the electrons, causing the synchrotron spectrum to
intensify and to extend farther into the submillimetre band.
Consequently, the Compton up-scattered X-ray emission would harden as
the X-ray intensity increased, exactly as observed.  We note that the
millimetre-band spectrum of \sgrastar\ has been observed to harden
during one 3-week flare\nolinebreak\cite{Tsuboi99} and one 3-day
flare\nolinebreak\cite{Wright93}, as would be required by the current
SSC models for \sgrastar\ (refs 20, 22).

To test the SSC models, we measured the flux density of \sgrastar\ at
a wavelength of 3~mm with the Millimeter Array at the Owens Valley
Radio Observatory, simultaneous with part of the 2000 \emph{Chandra}
measurement.  Unfortunately, the available observing window
(20:10--02:30 UT) preceded the X-ray flare (04:03--06:50 UT) by a few
hours.  The observed flux density of \sgrastar\ was $2.05 \pm 0.3$~Jy,
consistent with previously reported
measures\nolinebreak\cite{Serabyn97, Falcke98}.  Recently, a 106-day
quasi-periodicity has been reported in the centimetre band from an
analysis of 20 years of data taken with the Very Large Array
(VLA)\nolinebreak\cite{Zhao01}.  A weekly VLA monitoring program
detected an $\approx30$\% increase in the radio flux density of
\sgrastar\ beginning around 24 October 2000 and peaking on 5 November
2000.  This increase was seen at 2~cm, 1.3~cm, and 7~mm (R.~McGary,
J.-H.~Zhao, W.~M.~Goss, and G.~C.~Bower, private communication).  The
timing of the X-ray flare and the rise in the radio flux density of
\sgrastar\ suggests there is a connection between the two events,
providing additional indirect support for the association of the X-ray
flare with \sgrastar\ and further strengthening the case that it was
produced via either the SSC or direct synchrotron processes.
Definitive evidence for these ideas will require detection of
correlated variations in the radio-to-submillimetre and X-ray
wavebands through future coordinated monitoring projects.

}


\medskip

{\small\bf\noindent Acknowledgements}

\smallskip

{\footnotesize\noindent We thank M.~Begelman for useful comments.
This work has been supported by a grant from NASA.

\smallskip

\noindent Correspondence and requests for material should be addressed
to F.K.B. (e-mail: fkb@space.mit.edu).}


\begin{thebibliography}{}

\bibitem{Richstone98} Richstone, D.\ \emph{et~al.}  Supermassive black
holes and the evolution of galaxies.  \emph{Nature} {\bf 395}, A14--19
(1998).

\bibitem{Genzel00} Genzel, R., Pichon, C., Eckart, A., Gerhard, O.~E.\
\& Ott, T.\ Stellar dynamics in the Galactic Centre: proper motions
and anisotropy.  \emph{Mon.\ Not.\ R.\ Astron.\ Soc.}  {\bf 317},
348-374 (2000).

\bibitem{Ghez00} Ghez, A.~M., Morris, M., Becklin, E.~E., Tanner, A.,
\& Kremenek, T.\ The accelerations of stars orbiting the Milky Way's
central black hole.  \emph{Nature} {\bf 407}, 349--351 (2000).

\bibitem{Lynden-Bell71} Lynden-Bell, D.\ \& Rees, M.~J.\ On quasars,
dust and the Galactic Centre. \emph{Mon.\ Not.\ R.\ Astron.\ Soc.}
{\bf 152}, 461--475 (1971).

\bibitem{Melia01c} Melia, F.\ \& Falcke, H.\ The supermassive black
hole at the Galactic Center. \emph{Annu.\ Rev.\ Astron.\ Astrophys.}
{\bf 39}, (in the press).

\bibitem{Baganoff01} Baganoff, F.~K.\ \emph{et~al.} Chandra X-ray
spectroscopic imaging of \sgrastar\ and the central parsec of the
Galaxy. \emph{Astrophys.~J.} (submitted); also preprint
astro-ph/0102151 at $\langle$xxx.lanl.gov$\rangle$ (2001).

\bibitem{Morris96} Morris, M.\ \& Serabyn, E.\ The galactic center
environment.  \emph{Annu.\ Rev.\ Astron.\ Astrophys.} {\bf 34},
645--702 (1996).


\bibitem{Menten97} Menten, K.~M., Reid, M.~J., Eckart, A.\ \& Genzel,
R.\ The position of Sagittarius~A$^*$: accurate alignment of the radio
and infrared reference frames at the Galactic Center.
\emph{Astrophys.~J.} {\bf 475}, L111--L114 (1997).

\bibitem{Weisskopf96} Weisskopf, M.~C., O'Dell, S.~L.\ \& van
Speybroeck, L.~P.\ Advanced X-Ray Astrophysics Facility (AXAF).
\emph{Proc.\ SPIE} {\bf 2805}, 2--7 (1996).

\bibitem{Predehl95} Predehl, P.\ \& Schmitt, J.~H.~M.~M.\ X-raying the
interstellar medium: ROSAT observations of dust scattering halos.
\emph{Astron. Astrophys.} {\bf 293}, 889--905 (1995).

\bibitem{Mushotzky93} Mushotzky, R.~F., Done, C.\ \& Pounds, K.~A.\
X-ray spectra and time variability of active galactic nuclei.
\emph{Annu.\ Rev.\ Astron.\ Astrophys.} {\bf 31}, 717-761 (1993).

\bibitem{Davies98} Davies, M.~B., Blackwell, R., Bailey, V.~C.\ \&
Sigurdsson, S.\ The destructive effects of binary encounters on red
giants in the Galactic Centre.  \emph{Mon.\ Not.\ R.\ Astron.\ Soc.}
{\bf 301}, 745--753 (1998).

\bibitem{Nandra97} Nandra, K., George, I.~M., Mushotzky, R.~F.,
Turner, T.~J.\ \& Yaqoob, T.\ ASCA observations of Seyfert~1
galaxies. I.~data analysis, imaging, and timing.  \emph{Astrophys.~J.}
{\bf 476}, 70--82 (1997).

\bibitem{Ptak98} Ptak, A., Yaqoob, T., Mushotzky, R., Serlemitsos, P.\
\& Griffiths, R.\ X-ray variability as a probe of advection-dominated
accretion in low-luminosity active galactic nuclei.
\emph{Astrophys.~J.} {\bf 501}, L37--L40 (1998).

\bibitem{Ulrich97} Ulrich, M.-H., Maraschi, L.\ \& Urry, C.~M.\
Variability of active galactic nuclei.  \emph{Annu.\ Rev.\ Astron.\
Astrophys.} {\bf 35}, 445--502 (1997).

\bibitem{Narayan98} Narayan, R., Mahadevan, R., Grindlay, J.~E.,
Popham, R.~G.\ \& Gammie, C.\ Advection-dominated accretion model of
Sagittarius~A$^*$: evidence for a black hole at the Galactic center.
\emph{Astrophys.~J.} {\bf 492}, 554--568 (1998).

\bibitem{Quataert99} Quataert, E.\ \& Narayan, R.\ Spectral models of
advection-dominated accretion flows with winds.  \emph{Astrophys.~J.}
{\bf 520}, 298-315 (1999).

\bibitem{Ball01} Ball, G.~H., Narayan, R.\ \& Quataert, E.\ Spectral
models of convection-dominated accretion flows.  \emph{Astrophys.~J.}
{\bf 552}, 221--226 (2001).

\bibitem{Blandford99} Blandford, R.~D.\ \& Begelman, M.~C.\ On the
fate of gas accreting at a low rate on to a black hole.  \emph{Mon.\
Not.\ R.\ Astron.\ Soc.} {\bf 303}, L1--L5 (1999).

\bibitem{Falcke00} Falcke, H.\ \& Markoff, S.\ The jet model for
Sgr~A$^*$: radio and X-ray spectrum.  \emph{Astron.\ Astrophys.}  {\bf
362}, 113--118 (2000).

\bibitem{Melia94} Melia, F.\ An accreting black hole model for
Sagittarius~A$^*$. II: A detailed study.  \emph{Astrophys.~J.} {\bf
426}, 577--585 (1994).

\bibitem{Melia01a} Melia, F., Liu, S.\ \& Coker, R.\ A magnetic dynamo
origin for the submillimeter excess in Sagittarius~A$^*$.
\emph{Astrophys.~J.} {\bf 553}, 146--157 (2001).

\bibitem{Rees88} Rees, M.~J.\ Tidal disruption of stars by black holes
of $10^6$--$10^8$ solar masses in nearby galaxies.  \emph{Nature} {\bf
333}, 523--528 (1988).

\bibitem{Haardt97} Haardt, F., Maraschi, L.\ \& Ghisellini, G.\ X-ray
variability and correlations in the two-phase disk-corona model for
Seyfert galaxies.  \emph{Astrophys.~J.} {\bf 476}, 620--631 (1997).

\bibitem{Tsuboi99} Tsuboi, M., Miyazaki, A.\ \& Tsutsumi, T.\ Flare of
Sgr~A$^*$ at short millimeter wavelengths.  in \emph{The Central
Parsecs of the Galaxy} (ed.\ Falcke, H.\ \emph{et~al.}) {\bf 186},
105--112 (ASP Conf.\ Ser., Astron.\ Soc.\ Pac., San Francisco, 1999).

\bibitem{Wright93} Wright, M.~C.~H.\ \& Backer, D.~C.\ Flux density of
Sagittarius~A at $\lambda = 3$ millimeters.  \emph{Astrophys.~J.}
{\bf 417}, 560--564 (1993).

\bibitem{Serabyn97} Serabyn, E.\ \emph{et~al.} High frequency
measurements of the spectrum of SGR A$^*$.  \emph{Astrophys.~J.} {\bf
490}, L77--L81 (1997).

\bibitem{Falcke98} Falcke, H.\ \emph{et~al.} The simultaneous spectrum
of Sagittarius~A$^*$ from 20~centimeters to 1~millimeter and the
nature of the millimeter excess.  \emph{Astrophys.~J.} {\bf 499},
731--734 (1998).

\bibitem{Zhao01} Zhao, J.-H., Bower, G.~C.\ \& Goss, W.~M.\ Radio
variability of Sagittarius~A$^*$---a 106 day cycle.
\emph{Astrophys.~J.} {\bf 547}, L29--L32 (2001)

\bibitem{Reid93} Reid, M.~J.\ The distance to the center of the
Galaxy.  \emph{Annu.\ Rev.\ Astron.\ Astrophys.} {\bf 31}, 345--372
(1993).

\end{thebibliography}
\end{document}